\documentclass[aps,prl,twocolumn,superscriptaddress,groupedaddress,amsmath,amssymb]{revtex4}  
\usepackage{graphicx}
\usepackage{dcolumn}
\usepackage{bm}        
\usepackage{amssymb}   
\newcommand{\st}{\scriptsize}
\usepackage{bm}
\usepackage[usenames ,dvipsnames]{xcolor}
\usepackage[toc,page]{appendix} 
\usepackage{color}
\usepackage{url}
\usepackage{dsfont}        
\usepackage{slashed}
\usepackage{epsfig}
\usepackage{multirow,array}
\usepackage{float}
\usepackage{fancyhdr}
\usepackage{indentfirst}
\usepackage{cancel}
\usepackage{tabularx}
\usepackage{simplewick}
\usepackage{amsmath}
\usepackage[normalem]{ulem}
\usepackage{slashed}
\usepackage{upgreek}
\usepackage{xtab,afterpage,longtable}
\RequirePackage[colorlinks=true
  ,urlcolor=Green
  ,anchorcolor=blue
  ,citecolor=orange
  ,filecolor=blue
  ,linkcolor=blue
  ,menucolor=blue
  ,pagecolor=blue
  ,linktocpage=true
  ,pdfproducer=medialab
  ,pdfa=true
]{hyperref}

\newcommand{\hst}[1]{\hbox{\st{#1}}}

\def\beq{\begin{equation}}
\def\eeq{\end{equation}}
\def\bea{\begin{eqnarray}}
\def\eea{\end{eqnarray}}
\begin{document}

\title{Imprint of Early Dark Energy in Stochastic Gravitational Wave Background}
     
\author{Chia-Feng Chang}
\email{chiafeng.chang@email.ucr.edu}
\affiliation{Department of Physics and Astronomy, University of California, Riverside, CA 92521, USA}
\date{\today}

\begin{abstract}
Early dark energy that relieves Hubble tension leaves a fingerprint in the primordial stochastic gravitational wave (GW) background that originates from cosmic string network. The signal is not only detectable with future planned GW experiments, but also distinguishable from other astrophysical and cosmological signals in the GW frequency spectrum. We find that the cosmic string GW spectrum can probe other new physics that influence the universe in post-Big-Bang-Nucleosynthesis with mid-band GW detection, which extends GW cosmic archaeology search region.
\end{abstract}

\maketitle

\section{Introduction}



Gravitational waves (GWs) search technologies \cite{Abbott:2016blz,Abbott:2016nmj,Abbott:2017vtc,Abbott:2017gyy,Abbott:2017oio} have strong ongoing research interest. Implications from these developments focus not only on discovering new astrophysical objects \cite{TheLIGOScientific:2017qsa,GBM:2017lvd}, but also probing gravitation physics \cite{Lombriser:2016yzn,Lombriser:2015sxa,Creminelli:2017sry,Ezquiaga:2017ekz,Baker:2017hug,Sakstein:2017xjx} and non-standard cosmologies \cite{Chang:2021afa,Chang:2019mza,Caldwell:2018giq,Cui:2018rwi,Cui:2017ufi,Battye:1994au,Blasi:2020wpy,Gouttenoire:2019rtn}; and GW observations has been used to measure the current Hubble rate $H_0$ \cite{Virgo:2021bbr,Abbott:2017xzu,Mukherjee:2019qmm,Gayathri:2020fbl,Mukherjee:2020kki}, but not yet sufficiently precise to resolve the Hubble tension.

The tension is a Hubble rate discrepancy between local observations, such as supernovae signals \cite{Riess:2016jrr,Riess:2018byc,Breuval:2020trd,Riess:2020fzl,Soltis:2020gpl,Freedman:2020dne}, lensing time delays \cite{Bonvin:2016crt,Birrer:2018vtm,Denzel:2020zuq,Yang:2020eoh,Birrer:2020tax,Millon:2019slk,Baxter:2020qlr} and non-local searches, e.g.~the cosmic microwave background (CMB) \cite{Aghanim:2018eyx,Aylor:2017haa,Aiola:2020azj,Choi:2020ccd}. Local measurements confirm $H_0$ is approximately five sigma statistical significance above observation in the CMB with assuming standard cosmological model ($\Lambda$CDM). This strong disagreement has been closely examined using many statistical \cite{Feeney:2017sgx,Zhang:2017aqn,Cardona:2016ems,Bennett:2014tka} and measurement methods \cite{Pesce:2020xfe,Wong:2019kwg,Birrer:2020tax,Huang:2019yhh,Freedman:2020dne,Freedman:2019jwv}, and re-examining potential technical issues \cite{Benedict:2006cp,Humphreys:2013eja,Efstathiou:2013via,Rigault:2014kaa,Kenworthy:2019qwq,Spergel:2013rxa} in both local and CMB measurements (see reviews \cite{Knox:2019rjx,Verde:2019ivm,Wong:2019kwg,Aylor:2018drw,Vagnozzi:2019ezj}). Various evidence suggests the discrepancy arises due to a currently unknown physical phenomenon outside conventional $\Lambda$CDM predictions.


V.~Poulin \textit{et al.}~\cite{Poulin:2018cxd,Poulin:2018dzj,Smith:2019ihp} showed that early dark energy (EDE) behaves as a cosmological constant when comoving scaling factor $a(t)$ is smaller than critical $a_c \sim 10^{-4}$, and is then diluted faster or equal to radiation-like component to relieve the Hubble tension. EDE contributes up to $20\%$ energy density of the universe (model-dependent) at $a_c$, consequently accelerating the universe expansion and hence slightly delaying the universe entering the matter-domination era. This framework brings $H_0$ in estimated CMB to be consistent with local measurements and also with measured high and low redshifts \cite{Agrawal:2019lmo,Smith:2020rxx,Murgia:2020ryi,Lin:2020jcb,Vagnozzi:2021gjh,Ye:2020btb,Ye:2021iwa,Seto:2021xua,DiValentino:2021izs}.


We adopted the simplest EDE model, i.e., slow-rolling potential $V(\phi) \propto \phi^{2n}$ with scalar field $\phi$ \cite{Copeland:2006wr,Linder:2007wa,Chang:2019tvx}. Effective mass for $\phi$ is lighter than the Hubble rate on the early universe, and hence Hubble friction overdamps scalar field motion and freezes it. Consequently, the scalar field behaves as a subdominant cosmological constant until the Hubble rate decreases to approximately scalar effective mass, namely, the driving force overcomes the Hubble friction. Subsequently, the field starts oscillating as a fluid with equation of state $w_\phi = (n-1)/(n+1)$. As we will see that such an exotic cosmic component influences GW background that sourced by cosmic string, and leaves its fingerprint.

Cosmic strings, one dimension long-lived topological defects, are stable and predictable sources for stochastic GW background (SGWB) and hence ideal creator sources for GW as a messenger that carries new signal from the early universe \cite{Vilenkin:1984ib,Caldwell:1991jj,Hindmarsh:1994re}. Such stable objects arise from beyond standard model theories, such as spontaneously broken $U(1)$ \cite{Kibble:1976sj,Nielsen:1973cs,Vachaspati:1984dz,Vilenkin:2000jqa} or superstring theories \cite{Copeland:2003bj,Dvali:2003zj,Polchinski:2004ia,Jackson:2004zg}. The GW frequency spectrum formed by the cosmic string network is an approximate plateau over a broad frequency range with parameter $G \mu$ dependence, where $G$ is the Newtonian gravitational constant and $\mu$ is string tension. The plateau-like spectrum is an ideal tool to test new physics prior to Big Bang Nucleosynthesis (BBN) \cite{Cui:2017ufi,Cui:2018rwi,Gouttenoire:2019kij}, which have called \textit{cosmic archaeology}. However, we show that cosmic string SGWB can also test post-BBN physics with mid-band GW detectors, which extends the GW cosmic archaeology to CMB period.  GW experiments EPTA \cite{Lentati:2015qwp} and PPTA \cite{Shannon:2015ect} provide strong bounds on $G\mu \lesssim 2\times 10^{-11}$ \cite{Cui:2018rwi}, but this remains in tension with the NANOGrav 12.5yr result \cite{NANOGrav:2020bcs} $G\mu \in (2,30)\times 10^{-11}$ in 95\% C.L. \cite{Ellis:2020ena,Blasi:2020mfx}. The tension may be due to the different noise analysis methods \cite{Hazboun:2020kzd}. 

This paper shows that EDE imprints a GW signal in the SGWB formed from the cosmic string network that is distinguishable from other astrophysical and cosmological signals in the GW frequency spectrum. Such a unique spectrum could be detected in future mid-band GW experiments such as interferometer experiments DECIGO \cite{Seto:2001qf,Kawamura:2006up,Kawamura:2011zz,Sato:2017dkf,Kawamura:2020pcg}, BBO \cite{Yagi:2011wg,Crowder:2005nr,Corbin:2005ny,Harry:2006fi} and LISA \cite{LISA:2017pwj,Bartolo:2016ami,Caprini:2019pxz}, and pulsar timing observation SKA \cite{Janssen:2014dka,Carilli:2004nx,Weltman:2018zrl} for low frequency band. 





\section{Framework}

We consider the calculation method from an analytical derivation model, velocity-dependent one scale (VOS) model \cite{Martins:1995tg,Martins:1996jp,Martins:2000cs,Auclair:2019wcv}, with a calibration to simulation result from \cite{Blanco-Pillado:2013qja,Blanco-Pillado:2017oxo}. These strings formed at time $t_F$ when temperature cools to symmetry breaking scale of theory. Shortly after, the defects behave as a scaling invariant network that includes a few long (super-horizon length) strings and a collection of closed loops chopped from long strings. Loop number density in such a network can be characterized as \cite{Vilenkin:2000jqa}
\begin{align}
\label{Eq: n_o}
n_o(t_i,t) = \frac{0.1}{\alpha}  \int_{t_F}^t C_{\rm eff}(t_i) \frac{dt_i}{t_i^4} \left( \frac{a(t_i)}{a(t)} \right)^{3},
\end{align}
where loops form at $t_i$ and subsequently continuously diluted until time $t$, factor $0.1$ represents $90\%$ string energy release to loop kinetic energy and subsequent redshift away without transferring to GWs \cite{Vanchurin:2005pa,Olum:2006ix,Martins:2005es,Ringeval:2005kr,Blanco-Pillado:2011egf,Blanco-Pillado:2013qja,Blanco-Pillado:2017oxo}, and loop length parameter $\alpha=0.1$ \cite{Blanco-Pillado:2013qja,Blanco-Pillado:2017oxo}. The $t_i$ dependent function $C_{\rm eff}(t_i)$ varies with the cosmic equation of state and is sensitive to VOS calibration parameters, we leave a review for it in the Appendix. After $t_i$, loops start oscillating and emitting energy in the form of GWs with a constant rate
\begin{align}
\frac{dE}{dt} = - \Gamma G \mu^2,
\end{align}
where $E$ is loop energy, and GW emission parameter $\Gamma=50$ \cite{Blanco-Pillado:2013qja,Blanco-Pillado:2017oxo,Vilenkin:1981bx,Turok:1984cn,Quashnock:1990wv}. The loops consequently shorten loop length by radiating GWs as
\begin{align}
\ell(t) = \alpha t_i - \Gamma G \mu (t-t_i), \;\;\;\;\;\;\hbox{with}\;\;\;\; t\geq t_i,
\end{align}
where initial loop size $\ell(t_i)=\alpha t_i$. String loops emit GWs at time $t$ with frequency $f_{\rm emit} = 2k/\ell$ through normal model $k \in \mathbb{Z}^+$ oscillations. GWs redshifts as $a^{-1}$ until today ($t_0$). The observed GW frequency thus reads
\begin{align}
\label{Eq: frequency equal 2k ell}
f = \frac{2 k }{\ell} = \frac{a(t)}{a(t_0)} \frac{2k}{\alpha t_i - \Gamma G \mu (t-t_i)}.
\end{align}
With a collection of GWs that radiated by loops, the SGWB frequency spectrum can be computed as a summation of loops with all normal modes,
\begin{align}
\label{Eq: Omega_GW f 0}
\Omega_{\hst{GW}}(f) = \frac{f}{\rho_c}\frac{d \rho_{GW}}{df} = \sum_k \Omega_{\hbox{\st{GW}}}^{(k)}(f),
\end{align}
with
\begin{align}
\label{Eq: Omega_GW f}
 & \Omega_{\hst{GW}}^{(k)}(f) = \frac{1}{\rho_c} \int^{t_0}_{t_F} dt\, \Gamma^{(k)}G\mu^2 f \frac{dn_o(t_i,t)}{df} \left( \frac{a(t)}{a(t_0)} \right)^4  \\
 \notag & \;\;\;\;\;\;\;\;\;\;\; \;\;= \frac{0.1}{\rho_c}\frac{2k}{f} \frac{\Gamma^{(k)} G\mu^2}{\alpha(\alpha+\Gamma G \mu)} \\ \notag & \times \int_{t_F}^{t_0} dt \frac{C_{\hbox{\st{eff}}}(t_i)}{t_i^4}\left( \frac{a(t)}{a(t_0)} \right)^5\left( \frac{a(t_i)}{a(t)} \right)^3 \theta(t_i-t_F)\theta(\ell(t)),
\end{align}
where critical density $\rho_c = 3H_0^2/8\pi G$, cusp dominates GW emission $\Gamma^{(k)} = \Gamma /(3.6\, k^{4/3})$ \cite{Cui:2018rwi,Auclair:2019wcv}, and sum over $k$ modes up to $k\leq 10^{5}$. We add two Heaviside step functions $\theta(t_i-t_F)\theta(\ell(t))$ at the second line for ensuring the positive definition and energy conservation.   

\begin{figure}[t]
\includegraphics[width=0.48\textwidth]{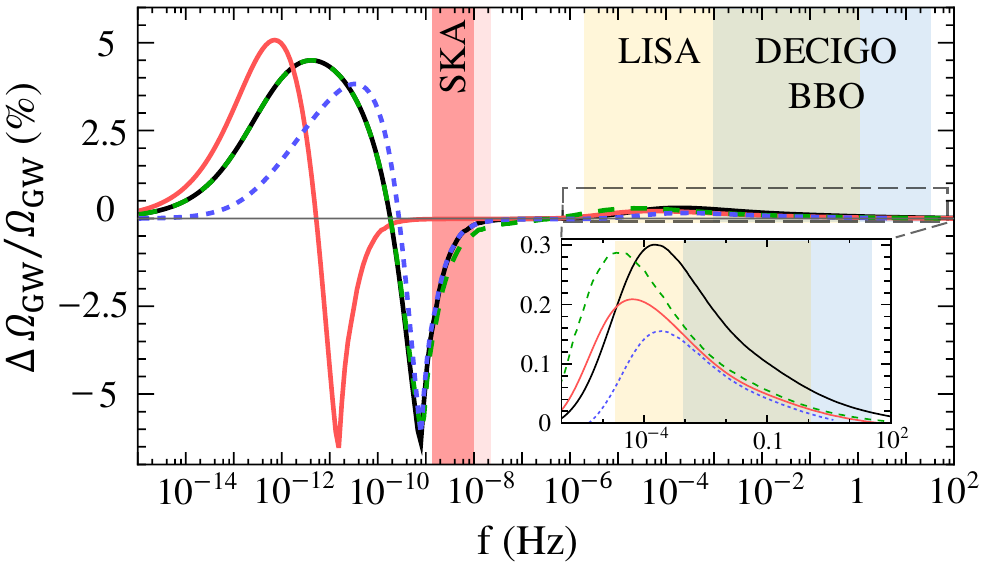} 
\caption{\label{Fig: Delta_Omega-ratio to f} Signal difference ratio versus GW frequencies as defined in Eq.(\ref{Eq: fluctuation to background ratio}). The black solid curve: $G\mu=10^{-12}$, $a_c = 10^{-4.48}$, $n=2$ and $f_{\rm EDE}=4\%$. According to black curve, others are changing one parameter on each, e.g.~red: change $a_c \to 10^{-3.57}$, green: $G\mu \to 10^{-11}$, and blue: $n \to 3$. The $a_c$ correspond to highest and lowest values in $68\%$C.L.~CMB analysis \cite{Poulin:2018cxd}. LISA and SKA sensitivities show as orange and red area, respectively. DECIGO and BBO sensitivities have presented as the blue area.}
\end{figure}

\begin{figure*}[t]
 \includegraphics[width=1.03\textwidth]{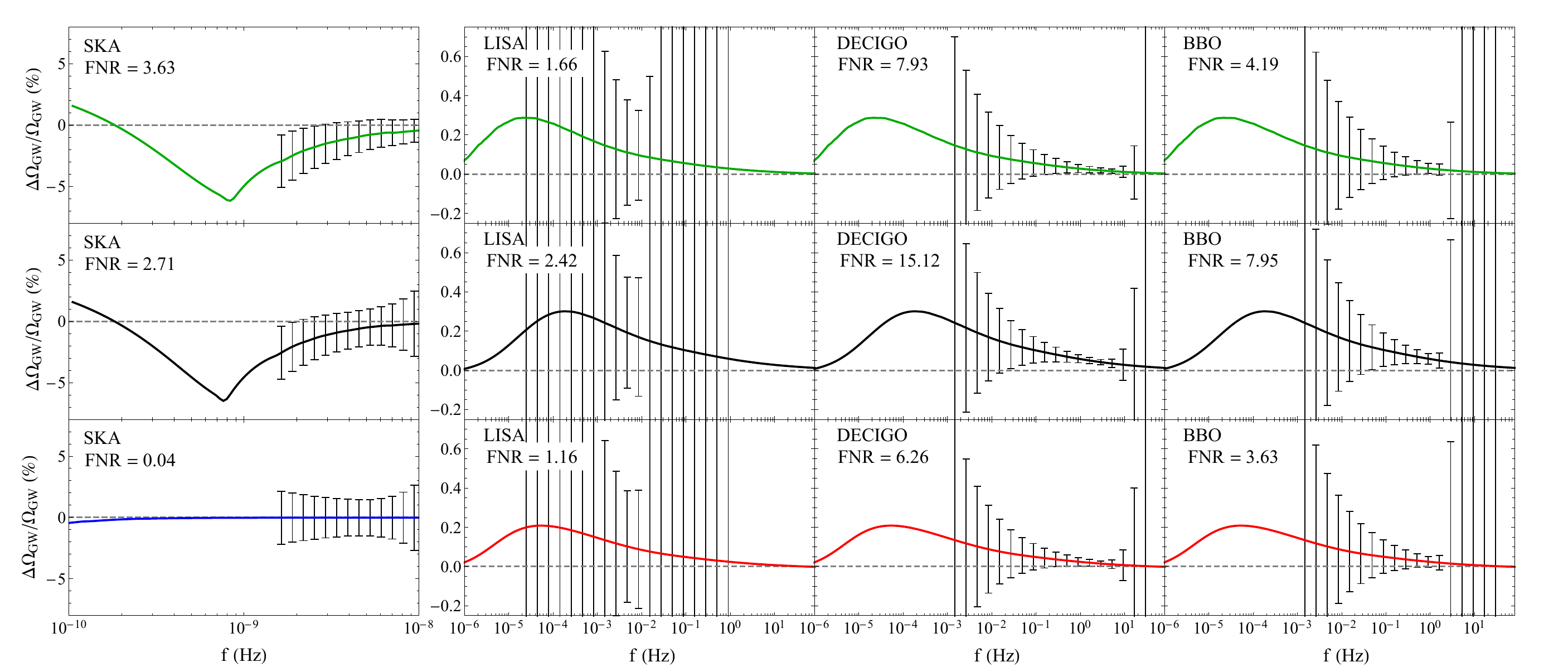}
  \caption{\label{Fig: Delta_Omega to fp} Signal difference ratio $\Delta \Omega_{\rm GW}/\Omega_{\rm GW}$ as GW frequency function with estimation of noise (including instrumental and self noise) as the error bars for 20 year SKA observation (with monitoring pulsars $N_{\rm SKA}=10^{3}$), 10 years LISA observation, and 5 years observation for upgraded-DECIGO, and BBO, respectively. We vary benchmark for the different row, and fix the early dark energy (EDE) total energy fraction $f_{\rm EDE} = 4\%$ and EDE potential $\phi^{2n}$ exponents $n = 2$, and varying string tension parameter $G\mu$ and critical scaling factor $a_c$ as shown in left plots. We also show the fluctuation-signal-to-noise ratio (FNR) as defined in Eq.(\ref{Eq: SNR}). }
\end{figure*}

To visualize EDE influence on SGWB, we define signal difference to SGWB spectrum ratio as,
\begin{align}
\label{Eq: fluctuation to background ratio}
\frac{\Delta \Omega_{\rm GW}}{\Omega_{\rm GW}}(f) \equiv \frac{\Omega_{\rm GW}^{\rm EDE}(f)-\Omega_{\rm GW}^{\st{\Lambda \hbox{CDM}}}(f)}{\Omega_{\rm GW}^{\st{\Lambda \hbox{CDM}}}(f)},
\end{align}
where superscripts imply GW frequency spectra on different cosmologies. 
Fig.~\ref{Fig: Delta_Omega-ratio to f} shows that EDE modifies the spectrum in two frequency regions: firstly, in contrast to $\Lambda$CDM, diluting EDE delays entering of matter domination, which slows down the redshift on GWs that emitted by the loops with loop formation time $t_i \lesssim \Gamma G \mu t_c/ \alpha$ where $t_c$ is the universe age at $a_c$. This mechanism \textit{peaks} the GW spectrum at frequency $f\sim 10^{-4}$~Hz. The resulting spectrum typically increases at characterized frequency $f_p$, which can be estimated from Eq.(\ref{Eq: frequency equal 2k ell}) as,
\begin{align}
f_{p} \sim  \frac{2\epsilon}{\Gamma G \mu t_c} \frac{a_c}{a(t_0)},
\end{align}
where $\epsilon \sim \mathcal{O}(1)$ is a numerical parameter. The shape of this peak approximately estimate to
\begin{align}
\label{Eq: Delta Omega over Omega f}
\frac{\Delta \Omega_{\rm GW}}{\Omega_{\rm GW}}(f) \propto \left\{            
\begin{aligned}
&\left( \frac{f}{f_{p}} \right)^{-0.23}, \;\;\; \;\;\;\;\; \hbox{for}\;\; f\geq f_p, \\
&\left( \frac{f}{f_{p}} \right)^{3/2} , \;\;\;\;\;\; \;\;\;\;\; \hbox{for}\;\; f < f_p,
\end{aligned}
\right.
\end{align}
and the slower diluting EDE (lower $n$) more significantly delays the universe entering the matter domination epoch while GWs experience a slower dilution effect, hence increasing the signal difference. 

Second, faster universe expansion at $t_c$ reduces loop chopping efficiency i.e., decreases $C_{\rm eff}(t_i=t_c)$ in Eq.(\ref{Eq: n_o}), due to less frequent intercommutation between strings. Such a mechanism implies a signal difference \textit{dip} at characteristic frequency $f_d \sim 10^{-9}$ to $10^{-11}\,$Hz as shown in Fig.~\ref{Fig: Delta_Omega-ratio to f}, which can be computed from Eq.(\ref{Eq: frequency equal 2k ell}) with GWs emission today,
\begin{align}
\label{Eq: f_d}
f_{d} \sim  \frac{2}{\alpha t_c}.
\end{align}
Shortly after $t_c$, more long strings are entering the horizon, and again, the diluting EDE slows down the universe expansion. Consequently, loop number density and $C_{\rm eff}$ increase slightly. This mechanism increases signal difference at frequencies below but close to $f_d$. 
 
Universal fitting on a wide frequency spectrum not only determines string tension parameter $G\mu$, but also addresses EDE parameters. $f_{\rm EDE}$ proportionally controls both signal difference peak and dip amplitudes, whereas peak amplitude is sensitive to $n$ but dip amplitude is independent. Therefore, we expect peak and dip amplitudes pin down $f_{\rm EDE}$ and $n$, and $f_d$ and $f_p$ can address $a_c$ (as shown in Fig.~\ref{Fig: Delta_Omega-ratio to f}). Those observations can also be complementary to other $H_0$ measurements.

In short, non-standard cosmology influences the cosmic string GW radiation mechanism at loop radiation and loop formation, which cause the peak and dip in GW spectrum, respectively. The former extends the GW cosmic archaeology to CMB period and opens a new window to test new post-BBN physics. 

Modified GW frequency spectra present in BBO/DECIGO, LISA, and SKA frequency sensitivities (Fig.~\ref{Fig: Delta_Omega-ratio to f}), and we focus on analyzing the detectability of the new signal on those experiment frequency bands for the remainder of this paper.

\section{Fluctuation-signal-to-Noise Ratio}

Early dark energy has relatively small energy fraction in the early universe, hence its influence GW variation is small compared to string SGWB, but such a fluctuation signal could still be detectable. In the rest of the paper, we assume that the future detectors can successfully perform source
separation of a detected SGWB signal (if exist other GW signal that comparable to or stronger than cosmic string one), and precisely detect the cosmic string GW. For comparing signal difference and noise spectrum, we use Fluctuation-signal-to-Noise Ratio (FNR) as $(i= {\rm BBO}, {\rm DECIGO}, {\rm LISA}, {\rm SKA})$,
\begin{align}
\label{Eq: SNR}
\hbox{FNR} = \sqrt{T_i \int^{f_{\rm max}}_{f_{\rm min}} df \left( \frac{\Delta\Omega_{\rm GW}(f)}{\Omega^{\rm eff}_{{\rm noise},i}(f)} \right)^2},
\end{align}
where experiments are calculated separately, frequency is integrated on the experimental sensitivity region, $T_i$ is observation period, and $\Omega^{\rm eff}_{{\rm noise},i}(f)$ is an effective GW noise spectrum including cosmic string SGWB self-noise and instrumental-noises for each experiment, respectively. We show the derivation of FNR and numerical details in the Appendix. 


Fig.~\ref{Fig: Delta_Omega to fp} shows signal difference ratio $\Delta\Omega_{\rm GW}/\Omega_{\rm GW}$ as a GW frequency function for each experiment sensitivity. We also mark the FNR and estimate the noise induced uncertainties for each benchmark with assumption that the signal and noise distribution are Gaussian as e.g.~\cite{Maggiore:1999vm}. We employ the experiment observation time $T$ as 5 years for BBO and upgraded-DECIGO, 10 years for LISA, and 20 years for SKA with monitoring pulsars $N_{\rm SKA}=10^3$ as a fiducial and assume that the pulsars are uniformly distributed in the sky. We use these assumptions in the rest of the paper.   

As shown, the signal difference distributes over a few frequency decades, including most sensitive experiment bands, see the gentle slope $f^{-0.23}$ over frequencies $f>f_p$ in Eq.(\ref{Eq: Delta Omega over Omega f}). Thus, the GW experiments can capture the EDE signal even though $f_p$ outsides their sensitivities. Higher $a_c$ corresponds to lower FNR because $f_p$ moves away from experiment sensitivities. 

\begin{figure}[t]
\flushleft\includegraphics[width=0.48\textwidth,trim=41 -5 42 -2, clip]{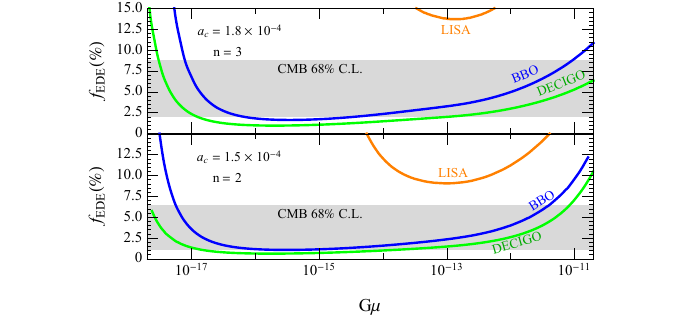}
\caption{\label{Fig: f_EDE vs Gmu} Energy density fraction of early dark energy $f_{\hst{EDE}}$ versus string tension parameter $G\mu$. The colored curves represent FNR$\,=5 $ with different experiments e.g.~BBO (blue), upgraded DECIGO (green), and LISA (orange). The EDE critical scaling factor $a_c$ and early dark energy potential exponent parameter $n$ have fixed on upper and lower plot, respectively. The gray areas present the CMB analysis with $68\% \,$C.L.~from \cite{Poulin:2018cxd}.}
\end{figure}


We explicitly show that the EDE signal could exceed the intensity of noise for interferometer experiments and SKA (if $N_{\rm SKA} \gtrsim 10^3$). The high FNR $\gtrsim \mathcal{O}(10)$ makes us more confident to detect such the sub-leading signal in SGWB. In addition, the self-noise dominates SKA uncertainty and makes EDE detection less likely. Increase $N_{\rm SKA}$ can proportionally decrease self-noise magnitude and thus increase the detectability, but $N_{\rm SKA}=10^3$ is already an optimistic assumption.


If the future detections can not perform a clean source separation for a detected SGWB signal, the signal difference can still be larger than observation uncertainties (FNR$>1$). For example, upgraded-DECIGO with benchmarks in Fig.~\ref{Fig: Delta_Omega to fp} and a strong astrophysical SGWB from mass binary black hole mergers with a merger rate of 56 $\hbox{yr}^{-1}\,\hbox{Gpc}^{-3}$ (see the spectrum in e.g.~Fig.~2 in \cite{Barish:2020vmy}) have FNR $\sim 2$ to $5$. Such a SGWB also contribute to the self-noise in Eq.(\ref{Eq: SNR}), and thus affects FNR.



The dependence on $G\mu$ for different $n$ shown in Fig.~\ref{Fig: f_EDE vs Gmu} matches our previous analysis. High FNR$\,>5$ can be obtained with wide range of string tension $ 10^{-17} \lesssim G\mu  \lesssim 10^{-11}$ for CMB $68\%$ C.L.~region for BBO/DECIGO detection. Decreasing string tension from $G\mu=10^{-11}$ would increase the peak frequency, which blue-shifts the EDE signal into experiment sensitivities. If a further decrease in the tension, the EDE signal would leave the search region and unlikely to detect, and the spectrum amplitude $\Omega_{\rm GW}(f)$ would be small for observation as well. Fig.~\ref{Fig: f_EDE vs Gmu} therefore shows a better detectability on mid $G\mu$ region. A faster EDE dispersion rate (i.e.~higher $n$) corresponds to a smaller cosmic expansion difference between EDE and $\Lambda$CDM, see Fig.~\ref{Fig: Delta_Omega-ratio to f}; in other words, lower $n$ implies better EDE detectability. 

\begin{figure}[t]
\includegraphics[width=0.48\textwidth,trim=0 0 0 5, clip]{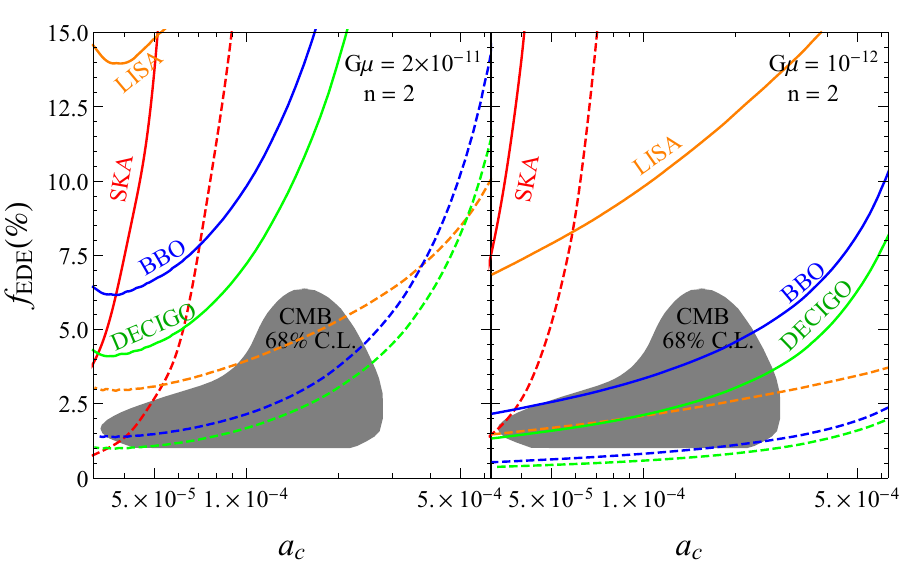}
\caption{\label{Fig: f_EDE vs ac} Energy density fraction of early dark energy $f_{\rm EDE}$ versus critical scaling factor $a_c$. The solid(dashed) curves represent FNR$\,=5$($=1$) for various experiments such as BBO (blue) and upgraded DECIGO (green), LISA (orange), and SKA (red), respectively. Our numerical uncertainties have shown as colored regions. The black areas present the CMB analysis with $68\% \,$C.L.~from \cite{Poulin:2018cxd}.}
\end{figure}

Fig.~\ref{Fig: f_EDE vs ac} shows experiment detection sensitivities to EDE parameter space. The experiments BBO and DECIGO would have a better accuracy for detecting the EDE signal. With the analysis in Fig.~\ref{Fig: f_EDE vs Gmu}, FNR$\,>5$ covers most CMB $68\%$ C.L.~region with our interested $G\mu$ range. Higher $a_c$ corresponds to lower FNR because $f_p$ moves away from sensitivities with increasing $a_c$. Changing of relativistic particle degree of freedom affects the $C_{\rm eff}$ at spectrum amplitude $\Omega_{\rm GW}(f)$ and induces a bump at $a_c \sim 3 \times 10^{-5}$ with $G\mu = 2 \times 10^{-11}$ as shown in left-panel.


If the NANOGrav signal is due to cosmic string i.e.~$G\mu \geq 2\times 10^{-11}$ \cite{Ellis:2020ena,Blasi:2020mfx}, the interferometer experiments should detect the EDE signal in future observations, and SKA has a possibility to capture the signal as shown in left-panel Fig.~\ref{Fig: f_EDE vs ac}. The statistical significance might not be large for individual experiments, but it is improvable with combinations of multiple experiments.


\section{Distinguish signal from other sources}

The EDE signal in cosmic string SGWB is distinguishable from other possible influences, such as string parameter variations, or sub-dominated SGWB from astrophysical or cosmological objects. For example, kinks or kink-kink collision modes \cite{Auclair:2019wcv}, string parameters $\alpha$ and $\Gamma$ variations \cite{Cui:2017ufi,Cui:2018rwi}, and new relativistic particle degrees of freedom in early universe \cite{Cui:2018rwi} would universally or uniformly influence GW amplitude. Therefore, they cannot cause local amplitude modification as does EDE. We numerically checked that EDE influenced SGWB difference is much flatter than cosmic string SGWB with influence from astrophysical objects, such as binary black hole merger \cite{Cholis:2016xvo,Mandic:2016lcn} and inspiral \cite{Bonetti:2020jku,Babak:2017tow,Amaro-Seoane:2007osp}. On the other hand, a sub-dominated GWs from cosmological phenomena, e.g. flat frequency spectrum from inflation \cite{Turner:1993vb,Boyle:2005se,Caprini:2018mtu}; peaky spectrum from domain wall \cite{Saikawa:2017hiv,Hiramatsu:2013qaa}; or first order phase transition dynamics \cite{Caprini:2009fx,Alanne:2019bsm,Schmitz:2020syl}: sound wave \cite{Hindmarsh:2015qta}, bubble wall collision \cite{Huber:2008hg,Weir:2016tov,Jinno:2016vai} and magnetohydrodynamic turbulence \cite{Caprini:2009yp,Binetruy:2012ze}, frequency spectra are distinguishable to EDE as well. In particular, the dip structure signal difference at $f_d$ cannot be caused by other known physical phenomena.

\section{Conclusion}

This paper proposed that SGWB originating from cosmic string network can be used to test the Hubble tension solution EDE. EDE behaves as dark energy in the early universe, then begins to dilute at the critical scaling factor $a_c\sim10^{-4}$ with total energy density fraction $f_{\rm EDE}\gtrsim 1\%$. It influences cosmic string SGWB by accelerating and decelerating universe expansion rate $\dot{a}(t)$ due to a dark energy like equation of state $\omega_{\phi}=-1$ and diluting faster than or equal to radiation-like component $\omega_\phi \geq 1/3$, respectively. The decelerated universe expansion rate locally increases cosmic string GW frequency spectra with magnitude $0.1\%$ to $1\%$ in the frequency range $10^{-5}$ to $10\,$Hz which is within interferometer experiment such as BBO/DECIGO and LISA sensitivity. And on the other side, the acceleration reduces spectra with magnitude $\sim 5\%$ in the lower frequency region $f \sim 10^{-9}$~Hz, within the SKA search region.  We also showed that the EDE influenced signals were stronger than the interferometer experiments and SKA (with $\gtrsim 10^3$ monitoring pulsars) noise spectrum, and hence detectable. Such spectral shapes are distinguishable from other SGWB sourced by cosmological or astrophysical objects. Thus, GW detection provides a possibility to probe EDE.

The introduced analysis method and new mechanism provide a new window for probing new physics influences in post-BBN cosmic history with mid-band GW detectors. This novel technique can test early physics not only for Hubble tension solutions, but also other physics that hide in cosmic history; in the words, it extends GW cosmic archaeology to CMB period. It can also be complementary to current early universe observations such as CMB and BBN. 




\acknowledgments

\section{Acknowledgement}

We thank Yanou Cui, Simeon Bird for valuable discussion. CC thanks Ming-Feng Ho for helpful comments on the draft. The author also thanks Academia Sinica for its hospitality. This project is in part supported by the Provost’s Scholars for the Advancement of Physical Sciences fellowship.

\appendix

\section{Appendix A: Velocity-dependent One-Scale Model}\label{A1}

\begin{figure*}[t]
\includegraphics[width=0.302\linewidth]{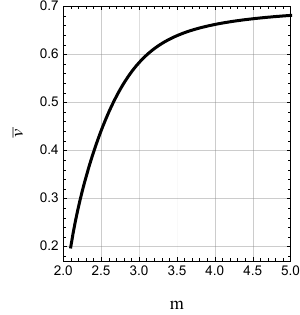}
\includegraphics[width=0.29\linewidth]{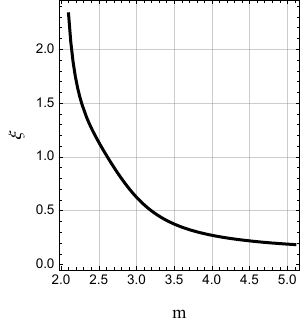}
\includegraphics[width=0.285\linewidth]{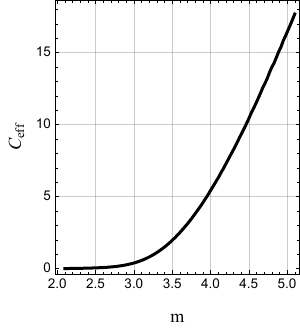}
\caption{\label{FVOS} Universe energy density $\rho \propto a^{-m}$ exponent dependence to VOS model parameters $\bar{v}$, $\xi$, and $C_{\rm eff}$, respectively.}
\end{figure*}

 In this section, we briefly review the velocity-dependent one-scale (VOS) model that used to predict the string network evolution \cite{Martins:1995tg,Martins:1996jp,Martins:2000cs}, and also reviewed in \cite{Cui:2018rwi,Cui:2017ufi,Avelino:2012qy,Sousa:2013aaa,Gouttenoire:2019kij}. 
 
 Cosmic strings have formed during a symmetry breaking phase transition when the vacuum manifold in the transition isn't simply connected. After such the transition, the cosmic strings rapidly enter a scaling regime. VOS model has successfully described the evolution of the cosmic string scaling network in terms of a mean string velocity
\begin{align}
\Bar{v} = \sqrt{\frac{m}{2} \frac{k(\bar{v})}{\left[ k(\bar{v})+\bar{c} \right]} \left(1-\frac{2}{m}\right)},
\end{align}
and a characteristic length,
\begin{align}
\xi = \frac{L}{t} = \frac{m}{2} \sqrt{\frac{k(\bar{v})\left[ k(\bar{v})+\bar{c} \right]}{2(m-2)}},
\end{align} as a fraction of the horizon,
where $L$ is long string length, $t$ is cosmic time, $m$ is exponent of scaling factor in universe energy density $\rho \propto a^{-m}$, the chopping parameter $\bar{c}=0.23$ \cite{Martins:2000cs}, and the ansatz function \cite{Martins:2000cs}
\begin{align}
k(\bar{v}) = \frac{2 \sqrt{2}}{\pi} (1-\bar{v}^2)\left( 1 + 2 \sqrt{2}\bar{v}^3\right) \frac{1-8\bar{v}^6}{1+8\bar{v}^6}.
\end{align}
The long string energy density express
\begin{align}
\rho_L = \frac{\mu}{\left(\xi t\right)^2}.
\end{align}
Those long strings lose energy through the intercommutation between them with creating loops as
\begin{align}
\frac{d\rho_L}{dt} = \bar{c}\bar{v} \frac{\rho_L}{\xi t} = \bar{c}\bar{v} \frac{\mu}{\left(\xi t\right)^3} .
\end{align}
Consequently, the loops number density at evolution time $t$ with particular loop size $\ell(t_i)=\alpha t_i$ that formed at time $t_i$ reads
\begin{align}
n_o(t_i,t) = \frac{0.1}{\alpha}  \int_{t_F}^t C_{\hbox{\st{eff}}}(t_i) \frac{dt_i}{t_i^4} \left( \frac{a(t_i)}{a(t)} \right)^{3},
\end{align}
with 
\begin{align}
\label{Eq: App: C_eff}
C_{\rm eff}(t_i) & = \frac{\bar{c}}{\gamma}\bar{v} \xi^{-3}, 
\end{align}
where $C_{\rm eff}$ implies that the loop energy gain from string network, $a^{3}$ is due to number density dilution, $\gamma=\sqrt{2}$ is loop Lorentz boost \cite{Blanco-Pillado:2013qja,Blanco-Pillado:2017oxo}. The $m$ dependence to VOS model parameters $\bar{v}$, $\xi$, and $C_{\rm eff}$ have shown in Fig.~\ref{FVOS}, respectively. 
EDE locally influences the universe expansion rate around $a(t)\sim a_c$, and therefore it would influence the $C_{\rm eff}$ on such a period. A smaller $C_{\rm eff}$ implies a faster universe expansion rate, i.e.~smaller $m$. It is because of a less frequent intercommuation rate in a faster-expanding universe.  

\begin{figure}[t]
\includegraphics[width=0.50\textwidth]{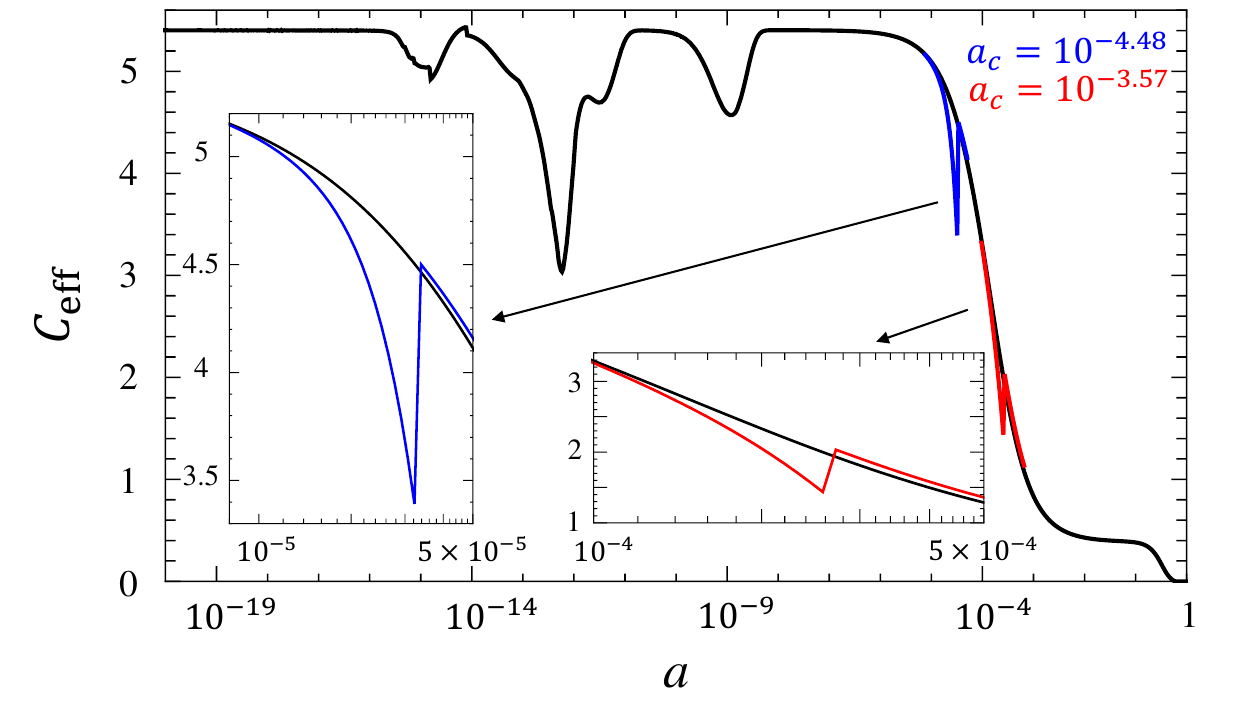}
\caption{\label{FCeff} $C_{\rm eff}$ versus scaling factor $a$. The black curve is in $\Lambda$CDM. The modified universes with the EDE are shown as $a_c=10^{-4.48}$ and $a_c=10^{-3.57}$ as blue and red, respectively.}
\end{figure}

We present $C_{\rm eff}$ versus scaling factor $a$ in Fig.~\ref{FCeff}, the changing relativistic degree of freedom causes the bumps on the black curve in the early universe. EDE locally influences $C_{\rm eff}$ around $a_c$ as the colored ranges. Before $a_c$, EDE behaves as a cosmological constant that speeds up the universe expansion rate, $C_{\rm eff}$ is thus smaller than the one in $\Lambda$CDM around the $a_c$. Suddenly, EDE dilutes at $a_c$ with a dilution rate that equal to or faster than a radiation-like components, and therefore, $C_{\rm eff}$ shows a sudden change. Then, the universe expansion rate $m$ turns to slightly larger than the $m$ in $\Lambda$CDM; $C_{\rm eff}$ increases to slightly higher. All curves converge when EDE energy density isn't comparable to other components. 

\section{Appendix B: Numerical Setup}\label{A2}

\subsection{Parameters}

The cosmological parameters we used in numerical are the scale factor for Hubble expansion rate $h=0.71$ \cite{Poulin:2018cxd}, pressureless matter density of the universe $\Omega_m = 0.31$, dark energy density of the $\Lambda$CDM universe $\Omega_\Lambda = 0.69$, today temperature $T_0=2.726\,$K, and relativistic degrees-of-freedom $g_*(T)$ from \cite{Saikawa:2018rcs}. For numerical uncertainty, we ignore extreme tiny difference $\Delta \Omega_{\rm GW}/\Omega_{\rm GW}(f) <0.001\%$ and $\Delta a(t)/a(t) < 0.01\%$. We have estimated the accuracy of $\Delta \Omega_{\rm GW}/\Omega_{\rm GW}(f)$ at about 10\% in our numerical, and this gives a good estimation for FNR. In addition, we expect that resolvable astronomical signals can be marginalized using frequency dependence (i.e.~wouldn't count them as noise, see for example \cite{Barish:2020vmy}) because, as discussed in the main letter, such signals have a distinct shape to our target.

\section{Appendix C: Interferometer and Pulsar Timing Experimental Noise Spectra}\label{A3}

We further discuss the noise analysis method for GW experiments, SKA, LISA, DECIGO, and BBO that we have considered in the main letter. We will derive the effective strain noise spectrum, estimate signal uncertainties, and introduce the fluctuation-to-noise ratio parameter in this section. In the first subsection Sec.~\ref{subsection: Formalism}, we introduce the formalism with following the analysis method developed in \cite{Allen:1996vm,Allen:1997ad,Maggiore:1999vm,Anholm:2008wy,Cutler:1997ta,Cornish:2001bb}, and reviewed in \cite{Schmitz:2020syl,Romano:2016dpx}. We apply the formalism to cross-correlated detectors in Sec.~\ref{subsection: Cross-Correlated Detectors}, and also introduce experiment parameters for DECIGO and BBO. Pulsar Timing Array and auto-correlated detector for LISA would be introduced in Sec.~\ref{subsection: PTA} and Sec.~\ref{subsection: ACD}, respectively.

\subsection{Formalism}\label{subsection: Formalism}

We focus on a stochastic, Gaussian, stationary, isotropic, and unpolarized GW background. We also adopt the conventions from Refs.\cite{Schmitz:2020syl} and \cite{Romano:2016dpx}. The goal of this subsection is to show an effective noise spectrum, then use it in the later subsections. 

For measuring the stochastic GW background, the observation data $S_I(t)$ is a combination of a signal and a noise for an auto-correlated detector (i.e.~single detector $I=1$) or cross-correlated detectors (multiple detectors $I=\{1,2,3,\dots\}$),
\begin{align}
\label{Eq: app: SI sI nI}
S_I(t) = s_I(t) + n_I(t),
\end{align}
where includes a signal contribution $s_I(t) = D^{ij}_I h_{ij}(t)$ with GW tensor perturbations $h_{ij}(t)$ at a given point $\vec{x}=0$ and detector tensor $D_{I}^{ij}$, and a noise contribution $n_I(t)$. 
The GW tensor perturbations in transverse traceless gauge can be decomposed into plane waves as
\begin{align}
h_{ij}(t) = \sum_{p = +, \times} \int_{-\infty}^{+\infty} df \int d^2 \textbf{n} \, h^p_{\textbf{n}}(f) (e^p_{\textbf{n}})_{ij} \, e^{2\pi i f (t-\textbf{nx})},
\end{align}
where $\textbf{n}$ is GW propagation direction, GW polarization $p$, $h^p_{\textbf{n}}(f)$ is an amplitude of a sinusoidal plane wave, and $(e^p_{\textbf{n}})_{ij}$ represents the corresponding polarization tensor
\begin{align}
\notag &(e^p_{\textbf{n}})_{ij} = (e^p_{\textbf{n}})_{ji}, \;\;\;\;\; (e^p_{\textbf{n}})_{ii} = 0, \;\;\;\;\;  \\ \notag & n_i (e^p_{\textbf{n}})_{ij} =0, \;\;\;\;\; 
(e^p_{\textbf{n}})_{ij} (e^{p^\prime}_{\textbf{n}})_{ij}^* = 2 \delta^{p p^\prime}.
\end{align}
Ensemble averaged (Gaussian expectation value) metric perturbation product gives 
\begin{align}
\label{Eq: app: hh}
\langle h^{p*}_{\textbf{n}}(f)h^{p^\prime}_{\textbf{n}^\prime}(f^\prime) \rangle = \frac{1}{4 \pi} \delta (f - f^\prime) \delta^{(2)}_{\textbf{n}\textbf{n}^\prime} \delta_{pp^\prime} \frac{1}{2} \frac{1}{2} S_{\rm signal}(f).
\end{align}
where the one of $1/2$ is due to the symmetry on frequency integration $h^{p*}_{\textbf{n}}(f) = h^{p}_{\textbf{n}}(-f)$, and the another is caused by the normalization of the polarization tensors. We assume SGWB to be unpolarizaed, isotropic, and stationary in Eq.(\ref{Eq: app: hh}), and the GW strain power spectrum $S_{\rm signal}(f)$ is a summation of polarization states, and integrated over the sky and presents as a function of frequency. The GW energy density spectral in such normalization is given \footnote{Note that a factor of $2$ difference on the definition e.g. in  Refs.\cite{Maggiore:1999vm,Caprini:2018mtu,Saikawa:2018rcs} is due to normalization of the polarization tensor (see also footnote 11 in \cite{Schmitz:2020syl}).},
\begin{align}
\label{Eq app Omega_gw}
\Omega_{\rm gw}(f) = \frac{2 \pi^2}{3 H_0^2} f^3 S_{\rm signal}(f),
\end{align}
where $H_0$ is current Hubble rate. We assume the signal and the noise modes are Gaussian as in \cite{Maggiore:1999vm}, and therefore, the expectation values for those modes are zero i.e.~$\langle s_I(t)\rangle = \langle n_I(t) \rangle = 0$. The quadratic expectation value for detector $I$ and $J$ reads,
\begin{align}
\label{Eq app Dnoise}
\langle \tilde{n}_I(f) \tilde{n}_{J}^*(f) \rangle = \frac{1}{2} \delta(f-f^\prime) \delta_{IJ} D_{\rm noise}^I (f),
\end{align}
where the tildes denote the Fourier transform, and $D^I_{\rm noise}(f)$ is the instrumental strain noise spectrum for detector $I$. For the multiple detector scenario, we assume that there is no noise correlation between detectors i.e.~$\langle n_I(f) n_{J}^*(f) \rangle = 0$ with $I \neq J$. The GW signal is translated by a response function $R^p_{\textbf{n},I}(f)$ that describes the antenna pattern of the detector as follows,
\begin{align}
\tilde{s}_I(f) = \sum_{p = +, \times} \int d^2 \textbf{n} \, R^p_{\textbf{n},I}(f) h_{\textbf{n}}^p(f).
\end{align}
The detail of the response function can be found in Refs.\cite{Maggiore:1999vm,Cornish:2001bb}. Then we can calculate the quadratic expectation value for the signal modes for a correlated detector pair $I$ and $J$,
\begin{align}
\label{Eq app sIsJ}
\langle \tilde{s}_I(f) \tilde{s}_J^*(f^\prime) \rangle = \frac{1}{2} \delta(f-f^\prime) \Gamma_{IJ}(f) S_{\rm signal}(f) ,
\end{align}
where we define a sky and polarization averaged overlap reduction function $\Gamma_{IJ}$,
\begin{align}
\label{Eq app GammaIJ}
    \Gamma_{IJ}(f) = \frac{1}{2} \sum_{ p = + , \times} \int \frac{d^2 \textbf{n}}{4 \pi} \, R^p_{\textbf{n},I}(f) R^{p*}_{\textbf{n},J}(f).
\end{align}
Furthermore, we also define a normalized overlap reduction function 
\begin{align}
    \gamma_{IJ}(f) = \frac{5}{\sin^2\delta} \Gamma_{IJ}(f),
\end{align}
where $\gamma_{IJ}(f=0)=1$, and $\delta = \pi/3$ is the opening angle between two arms of interferometers for LISA, DECIGO and BBO. We further define an effective instrumental strain noise power spectrum,
\begin{align}
\label{Eq: app Snoise}
    S_{\rm noise}^{\rm ins}(f) = \left( \sum_{J>I} \frac{\Gamma_{IJ}^2(f)}{D^I_{\rm noise}(f) D^J_{\rm noise}(f)} \right)^{-1/2},
\end{align}
where we only include the instrumental noise. We will use it in the next section. Then, similar to Eq.(\ref{Eq app Omega_gw}), we obtain the effective instrumental noise spectrum,
\begin{align}
\label{Eq: app: Omega_noise}
\Omega_{\rm noise}^{\rm ins}(f) = \frac{2 \pi^2}{3 H_0^2} f^3 S_{\rm noise}^{\rm ins}(f).
\end{align}
We will see an effective GW noise spectrum $\Omega^{\rm eff}_{\rm noise}(f)$ that includes both self-noise and instrumental noise in the next section. We further discuss cross-correlated (DECIGO and BBO), pulsar timing array (SKA), and auto-correlated (LISA) detector noise spectra in the following sections, respectively.

\subsection{Cross-Correlated Detectors}\label{subsection: Cross-Correlated Detectors}

We mainly follow the calculations in \cite{Maggiore:1999vm,Saikawa:2018rcs,Cornish:2001bb}, and now considering a cross-correlation measurement. Continuing the calculations from the last subsection, a measured cross-correlation signal $S_{IJ}$ from two detectors $I$ and $J$ ($I \neq J$) can be constructed with a filter function $Q_{IJ}(t - t^\prime)$,
\begin{align}
    S_{IJ} = \int^{t_{\rm obs}/2}_{-t_{\rm obs}/2} dt \int^{{t_{\rm obs}/2}}_{-{t_{\rm obs}/2}} dt^\prime S_I(t) Q_{IJ}(t - t^\prime) S_J(t^\prime).
\end{align}
The filter function $Q_{IJ}(t-t^\prime)$ falls rapidly to zero for large $t-t^\prime$, and we assume the observation period $t_{\rm obs}$ is much longer than the fall off timescale in $Q_{IJ}(t-t^\prime)$, and therefore, $S_{IJ}$ can be rewritten in frequency domain as
\begin{align}
    S_{IJ} = \int_{-\infty}^{+\infty} df \, \tilde{S}_I^*(f) \tilde{S}_J(f) \tilde{Q}(f).
\end{align}
Apply to the cross-correlated signal with coherent frequency, we have
\begin{align}
\label{Eq: app: s_IJ equiv}
\notag & \langle s_{IJ} \rangle \equiv \int^{+\infty}_{-\infty} df \langle \tilde{s}_I(f) \tilde{s}_J^*(f) \rangle \tilde{Q}(f) \\ & = \frac{t_{\rm obs}}{2} \int_{-\infty}^{+\infty} df \, \Gamma_{IJ}(f) S_{\rm signal}(f),
\end{align}
where we used Eq.(\ref{Eq app sIsJ}) and
\begin{align}
\notag \lim_{f^\prime \to f} \delta(f-f^\prime) & = \lim_{f^\prime \to f} \int^{t_{\rm obs}/2}_{-t_{\rm obs}/2} dt \, \exp\left[ - 2 \pi (f-f^\prime)t \right] \\ & = \lim_{f^\prime \to f} \frac{\sin\left[\pi (f-f^\prime) t_{\rm obs} \right]}{\pi (f-f^\prime)} = t_{\rm obs}.
\end{align}
The mean $\mu$ in such the cross-correlated detectors is given
\begin{align}
\label{Eq app mu}
\mu \equiv \langle S_{IJ} \rangle = \langle s_{IJ} \rangle = t_{\rm obs} \int_{0}^{\infty} df \, \Gamma_{IJ}(f) S_{\rm signal}(f),
\end{align}
where we employed the fact that the noise in two detectors aren't correlated. Then, the variation of signal can be found as
\begin{align}
\label{Eq app sigma}
\notag \sigma^2 \equiv & \, \langle S_{IJ}^2 \rangle - \langle S_{IJ}\rangle^2 \\
\notag =& \, \int_{-\infty}^{\infty} df\, df^\prime \tilde{Q}(f) \tilde{Q}^*(f^\prime) \Bigg[ \langle \tilde{S}^*_I(f) \tilde{S}_{J}(f) \tilde{S}_I (f^\prime) \tilde{S}^*_J(f^\prime) \rangle \\ \notag & - \langle \tilde{S}^*_I(f) \tilde{S}_J(f)\rangle \langle \tilde{S}_J^*(f^\prime) \tilde{S}_I (f^\prime) \rangle  \Bigg]\\
 \notag   =& \int_{-\infty}^{+\infty} df df^\prime\, \tilde{Q}(f)\tilde{Q}^*(f^\prime) \Bigg[ \langle \tilde{S}^*_I(f) \tilde{S}_I(f^\prime)  \rangle \langle \tilde{S}^*_J(f) \tilde{S}_J(f^\prime)  \rangle \\ \notag &  + \langle \tilde{S}^*_I(f) \tilde{S}^*_J(f^\prime)  \rangle \langle \tilde{S}_J(f) \tilde{S}_I(f^\prime)  \rangle  \Bigg] \\
\equiv & \, \frac{ t_{\rm obs}}{2} \int^\infty_0 df\, |\tilde{Q}(f)|^2 A(f),
\end{align}
where we define
\begin{align}
\label{Eq: app: A(f)}
    \notag A(f) \equiv & \left[ \Gamma_{I}(f)\Gamma_{J}(f) + \Gamma_{IJ}^2(f) \right] S_{\rm signal}^2(f)  \\ \notag &  + \left[ D_{\rm noise}^I(f)\Gamma_I(f) + D_{\rm noise}^J(f)\Gamma_J(f) \right]S_{\rm signal}(f)  \\  &   + D_{\rm noise}^I(f) D_{\rm noise}^J(f),
\end{align}
and used $S_{\rm signal}(f)=S_{\rm signal}(-f)$, $D_{\rm noise}^I(f)=D_{\rm noise}^I(-f)$, Eqs.(\ref{Eq: app: SI sI nI},\ref{Eq app Dnoise},\ref{Eq app sIsJ},\ref{Eq: app: s_IJ equiv}), and the Wick's theorem. Assuming two detectors are exactly identical, co-located, and co-aligned, we have
\begin{align}
   \notag  Q(t-t^\prime) \simeq \delta(t-t^\prime) \;\;\;\;\;\;\;\; \to \;\;\;\;\;\;\;\; \tilde{Q}(f) = 1,
\end{align}
the response function for a single detector, and the instrumental strain noise spectrum 
\begin{align}
\label{Eq: app: Gamma and D_noise}
\notag & \Gamma_I (f) = \Gamma_{II} (f) = \Gamma_{JJ} (f) \equiv \mathcal{R}_I(f), \;\;\;\;\;\; \\  & D^I_{\rm noise}(f) = D^J_{\rm noise}(f) \equiv D_{\rm noise}(f).
\end{align}
The mean and variance present with a discrete set of frequency bins $\{f_i\}$ with an integer $i$. According to Eq.(\ref{Eq app mu}) and Eq.(\ref{Eq app sigma}), the mean and variance at $i$-th bin are given
\begin{align}
\label{Eq: app: mu fi}
    \mu(f_i) =&\, t_{\rm obs} \Gamma_{IJ}(f_i)\, \Delta f_i\, \bar{S}_{\rm signal}(f_i),\\
\label{Eq: app: sigma fi}
    \sigma^2(f_i) =&\, \frac{t_{\rm obs}}{2}\, \Delta f_i\, \bar{A}(f_i),
\end{align}
where
\begin{align}
     \bar{S}_{\rm signal}(f_i) \equiv&\, \frac{1}{\Delta f_i} \int_{F_i} df\,  S_{\rm signal}(f),\\
    \bar{A}(f_i) \equiv&\, \frac{1}{\Delta f_i} \int_{F_i} df\, A(f),
\end{align}
with the integration rage $F_i = \left[ f_i - \Delta f_i/2, f_i + \Delta f_i/2 \right]$ for each frequency bin. Then use Eq.(\ref{Eq app Omega_gw}) and Eq.(\ref{Eq: app: mu fi}), we obtain the corresponding uncertainty of the spectrum $\Omega_{\rm gw}(f)$,
\begin{align}
\label{Eq: app: Delta Omega gw fi}
\notag \Delta \Omega_{\rm gw} (f_i) & = \frac{2\pi^2}{3H_0^2} \frac{f_i^3}{\Gamma_{IJ}(f_i) t_{\rm obs} \Delta f_i} \sigma(f_i) \\ & = \frac{2\pi^2}{3H_0^2} \frac{f_i^3}{\Gamma_{IJ}(f_i) \sqrt{2 t_{\rm obs} \Delta f_i}} \sqrt{\bar{A}(f_i)}.
\end{align}
In our numerical setup, we adopt the frequency resolution $\Delta f_i = f_i/10$ as in, for example, Refs.\cite{Cornish:2001bb,Saikawa:2018rcs}. Moreover, it is straightforward to see a comparison between the GW density spectrum uncertainty $\Delta\Omega_{\rm gw}(f)$ and a small deviation (fluctuation) between the GW density spectrum that predicted with non-standard and standard cosmology,
\begin{align}
\label{Eq: app: Omega th}
    \Delta \Omega_{\rm gw}^{\rm th}(f) \equiv \Omega_{\rm gw}^{\rm non-st}(f) - \Omega_{\rm gw}^{\rm st}(f).
\end{align}
Similar to Eq.(\ref{Eq: app: Omega th}), the mean difference between two theories:
\begin{align}
\label{Eq: app: delta mu}
    \Delta \mu(f_i) \equiv  t_{\rm obs} \Gamma_{IJ}(f_i)\, \Delta f_i\, \left[\bar{S}_{\rm signal}^{\rm non-st}(f_i) - \bar{S}_{\rm signal}^{\rm st}(f_i) \right].
\end{align}
In our analysis, we claim that the small deviation is detectable if 
\begin{align}
    \sum_{f_i} \left(\frac{\Delta \Omega_{\rm gw}^{\rm th}(f_i)}{\Delta \Omega_{\rm gw} (f_i) } \right)^2 \geq 1,
\end{align}
one can obtain the standard signal-to-noise ratio if we simply replace the numerator to the GW density spectrum mean, see e.g.~\cite{Cutler:1997ta,Maggiore:1999vm}. Then switch to an integration form with sum over all the frequencies $f_i$, we have
\begin{align}
\label{Eq: app: sum to inte}
\sum_{f_i} \left(\frac{\Delta \Omega_{\rm gw}^{\rm th}(f_i)}{\Delta \Omega_{\rm gw} (f_i) } \right)^2 =\int \left( \frac{\Delta \Omega_{\rm gw}^{\rm th}(f)}{\Omega^{\rm eff}_{\rm noise}(f)}\right)^2 \, 2 t_{\rm obs} \, df, \, 
\end{align}
where we define an effective GW noise spectrum,
\begin{align}
\label{Eq: app: effective Omega noise}
    \Omega^{\rm eff}_{\rm noise}(f) \equiv \frac{2\pi^2}{3H_0^2} f^3 \sqrt{ \frac{A(f)}{\Gamma_{IJ}^2(f)}} \simeq \Omega_{\rm noise}^{\rm ins}(f),
\end{align}
where $ \Omega_{\rm noise}^{\rm ins}(f)$ is given in Eq.(\ref{Eq: app: Omega_noise}). The approximation is only if the self-noise i.e.~the $S_{\rm noise}(f)$ terms in Eq.(\ref{Eq: app: A(f)}) are ignorable. This satisfies most cases, and has been applied in the literature e.g.~\cite{Schmitz:2020syl,Cutler:1997ta,Maggiore:1999vm}. But we include both self noise and instrumental noise in our calculation. To mimic the standard procedure for signal-to-noise ratio, we define a norm parameter "fluctuation-to-noise" ratio (FNR) that implies the detectability of $\Delta \Omega_{\rm gw}^{\rm th}(f)$:
\begin{align}
\label{Eq: app: FNR}
\hbox{FNR} \equiv \sqrt{ \left(\frac{\Delta \mu}{\sigma}\right)^2} =  \left[ 2\, t_{\rm obs} \int df\, \left( \frac{\Delta \Omega_{\rm gw}^{\rm th}(f)}{\Omega^{\rm eff}_{\rm noise}(f)}\right)^2\right]^{1/2},
\end{align}
where we used Eq.(\ref{Eq: app: sigma fi}) and Eq.(\ref{Eq: app: delta mu}) with summation of frequencies as Eq.(\ref{Eq: app: sum to inte}), and the factor $2$ implies a correlated detector pair. We will see an example with multiple detectors or an auto-correlated detector in the following sections. We claim that the deviation signal is detectable if $\hbox{FNR}\geq 1$. The observation period $t_{\rm obs}$ for cross-correlated interferometer experiments (DECIGO and BBO) is 5 years in our numerical setup. We will introduce other detector design-dependent functions, e.g. $\mathcal{R}_I(f)$, $\Gamma_{IJ}(f)$, and $D_{\rm noise}(f)$ for different experiments, respectively, in the later sections.

\subsubsection{upgraded DECIGO parameters}

Deci-hertz Interferometer Gravitational wave Observatory (DECIGO) is a planned Japanese space gravitational wave antenna \cite{Seto:2001qf,Kawamura:2006up,Kawamura:2011zz,Kawamura:2020pcg}.  DECIGO is targeted to observe both astrophysical and cosmological GWs for frequency bands 0.1 to 10 Hz. Such a sensitivity band is right above an irresolvable GW noise from many compact binaries that potentially confuse limiting noise level, see e.g.~\cite{Farmer:2003pa}. Also, DECIGO contains three independent data channels that could cross correlate to each other, which could exclude exact orthogonal noise, see Eq.(\ref{Eq app Dnoise}).  DECIGO thus potentially touches the extremely deep window in this band.

DECIGO has been designed for three drag-free spacecraft with satellite-borne triangular interferometers with opening angle $\delta = \pi/3$ in a hexagonal configuration. DECIGO overlap reduction functions has been computed in Refs.~\cite{Kudoh:2005as,Kuroyanagi:2014qza}, and more recently reviewed in Refs.~\cite{Schmitz:2020syl}. The numerical result from \cite{Kuroyanagi:2014qza} has shown in \cite{Schmitz:2020syl} Fig.11, we simply employ a good approximation on normalized overlap reduction function $\gamma_{IJ}(f)=1$ in low frequencies $f \leq 50\,$Hz, and rapidly fall off on higher frequencies $f>50\,$Hz, i.e.~
\begin{align}
    \Gamma_{IJ}(f)  =\left\{
\begin{aligned}
2  \frac{1}{5}\sin^2\beta \; & \gamma_{IJ}(f) \simeq \frac{3}{10}, \;\;\;\;\;\;\hbox{for}\;\;\;\; f\leq 50\,\hbox{Hz} \\
& 0, \;\;\;\;\;\;\;\;\;\;\;\;\;\;\;\;\;\;\;\;\;\;\;\;\hbox{for}\;\;\;\; f>50\,\hbox{Hz} 
\end{aligned}
\right.
\end{align}
where definition of $\beta$ is given in \cite{Nishizawa:2009bf}. On the other hand, the response function $\Gamma_{II}(f) \equiv \mathcal{R}_I(f)$ for a single equal-arm Michelson interferometer has analyzed in Ref.\cite{Larson:1999we}, its numerical result was computed in Fig.~10 of \cite{Schmitz:2020syl}, and given a closed analytic form as
\begin{align}
\label{Eq: app: signle overlap reduction function}
    \mathcal{R}_I(f) \simeq \frac{1}{5} \sin^2\delta \, \gamma_I(f),
\end{align}
with an approximation
\begin{align}
\gamma_I(f) \simeq \frac{1}{1+0.54(f/f_*^{\rm DECIGO})^2}.
\end{align}
The instrumental strain noise spectrum of DECIGO has been analyzed in \cite{Kuroyanagi:2014qza}. Three contributions dominate the instrumental strain noise spectrum $D_{\rm noise}(f)$,
\begin{align}
\notag D^{\rm DECIGO}_{\rm noise} (f) = & D_{\rm shot}^{\rm DECIGO}(f) \\ & + D_{\rm rad}^{\rm DECIGO}(f) + D_{\rm acc}^{\rm DECIGO}(f),
\end{align}
which are quantify shot noise, radiation pressure noise, and acceleration noise, respectively. They present as function of experiment designing parameters (we use the notation from \cite{Schmitz:2020syl})
\begin{align*}
D_{\rm shot}^{\rm DECIGO}(f) = &\, \frac{\hbar c_{\rm light} \pi \lambda}{P_{\rm eff}} \left( \frac{1}{4 F L_{\rm arm}^{\rm DECIGO}} \right)^2 \\ & \times \left[ 1 + \left( \frac{f}{f_{*}^{\rm DECIGO}} \right)^2 \right],\\
D_{\rm rad}^{\rm DECIGO}(f) = &\, \frac{\hbar P}{c_{\rm light} \pi \lambda} \left( \frac{16 F^{\rm DECIGO}}{M L_{\rm arm}^{\rm DECIGO}} \right)^2 \\ & \times \left( \frac{1}{2\pi f}\right)^4 \left[ 1 + \left( \frac{f}{f^{\rm DECIGO}_{*}}\right)^2 \right]^{-1},\\
D_{\rm acc}^{\rm DECIGO}(f) = &\, \frac{\hbar P}{c_{\rm light} \pi \lambda} \left( \frac{16 F^{\rm DECIGO}}{3M L_{\rm arm}^{\rm DECIGO}} \right)^2 \left( \frac{1}{2\pi f}\right)^4,
\end{align*}
where $\hbar$ is the Planck constant, $c_{\rm light}$ is the light speed, and DECIGO characteristic frequency $f_*^{\rm DECIGO} = c_{\rm light}/(2\pi L_{\rm arm}^{\rm DECIGO})$ \cite{Schmitz:2020syl}. We adopt the upgraded DECIGO parameters \cite{Kuroyanagi:2014qza,Kuroyanagi:2014qaa,Yagi:2011wg} as given: the laser output power $P = 30\,$W with wavelength $\lambda = 532\,$nm, the DECIGO arm length $L_{\rm arm}^{\rm DECIGO} = 1500\,$km, and the mirror mass $M=100\,$kg with radius $R = 0.75\,$m. The cavity parameter $F^{\rm DECIGO}$ is given
\begin{align}
F^{\rm DECIGO} = \frac{\pi (r_E r_F)^{1/2}}{1-r_E r_F}.
\end{align}
The effective laser output power reads,
\begin{align}
    P_{\rm eff} = \left( \frac{r_E t_F^2}{1-r_E r_F} \right)^2 P,
\end{align}
where the $r$'s parameters are calculated as 
\begin{align}
\notag   &r_E = r_{Em} r_G, \;\;\; r_F = r_{Fm} r_G, \;\;\; t_F =\sqrt{r_G^2 - r_{Fm}^2},\\
  \notag  &r_G = 1 - \hbox{exp}\left( - \frac{2 \pi R^2}{\lambda L_{\rm arm}^{\rm DECIGO}} \right), \;\;\;\; \\ & r_{Em}^2 = 0.9999, \;\;\;\; r_{Fm}^2 = 0.67.
\end{align}

\subsubsection{BBO parameters}

The Big Bang Observer (BBO) \cite{Corbin:2005ny,Crowder:2005nr,Harry:2006fi,Yagi:2011wg} has been proposed as a fleet of triangular interferometers operating on the out space. The GW observation principle is the same as DECIGO with the same frequency-sensitive band. It has been designed with three independent data channels, and they could cross correlate to each other. Therefore, BBO can search weak signals such as inflation sourced GWs \cite{Corbin:2005ny}.

The overlap reduction function for BBO has been calculated in Ref.\cite{Thrane:2013oya}, we quote the normalized overlap reduction function $\gamma_{IJ}(f)$ from Fig.5-right of \cite{Thrane:2013oya}, their numerical result can be downloaded in \footnote{J. Romano and E. Thrane, “Sensitivity curves for searches for gravitational-wave
backgrounds.” \url{https://dcc.ligo.org/LIGO-P1300115/public.}}, and the overlap reduction function reads
\begin{align}
\Gamma_{IJ}(f) = \frac{1}{5} \sin^2\delta \,\gamma_{IJ}(f) = \frac{3}{20}\, \gamma_{IJ}(f).
\end{align}
where we assume a single data channel as in \cite{Thrane:2013oya}. The response function for a single detector is given 
\begin{align}
\label{Eq: app: signle overlap reduction function 2}
    \mathcal{R}_I(f) = \Gamma_{I}(f) \simeq \frac{3}{20} \gamma_I(f) ,
\end{align}
where a numerical solution for $\gamma_I(f)$ is given in Fig.4 \cite{Thrane:2013oya}, and $f_*^{\rm BBO} = c_{\rm light}/(2\pi L_{\rm arm}^{\rm BBO})$. The BBO instrumental strain noise spectrum is given in \cite{Crowder:2005nr,Cutler:2005qq,Thrane:2013oya} 
\begin{align}
D_{\rm noise}^{\rm BBO} (f) = \frac{4}{(L_{\rm arm}^{\rm BBO})^2} \left[ D_{\rm oms}^{\rm BBO}(f) + \frac{1}{(2\pi f)^4} D_{\rm acc}^{\rm BBO}(f) \right],
\end{align}
where the position noise $D_{\rm oms}^{\rm BBO}(f)$ and acceleration noise $D_{\rm acc}^{\rm BBO}(f)$ are given, respectively \cite{Schmitz:2020syl},
\begin{align}
    \notag & D_{\rm oms}^{\rm BBO}(f) \simeq (1.4\times 10^{-17} \,\hbox{m})^2\,\hbox{Hz}^{-1}, \;\;\;\;\;\; \\ &  D_{\rm acc}^{\rm BBO}(f) \simeq (3\times 10^{-17} \,\hbox{m}\,\hbox{s}^{-2})^2\,\hbox{Hz}^{-1}.
\end{align}
The BBO arm is $L_{\rm arm}^{\rm BBO} = 5\times 10^{7}\,$m.

\subsection{Pulsar Timing Array}\label{subsection: PTA}


Pulsar timing arrays are monitoring an array of pulsars for timing residuals with quadrupole correlation i.e.~the Hellings-Downs curve. Increase regular observations of pulsar number can significantly enhance the pulsar timing precision \cite{Janssen:2014dka}, as we will discuss as follows. We follow the calculations that have been studied in \cite{Anholm:2008wy}, see also \cite{Hazboun:2019vhv,Chamberlin:2014ria,vanHaasteren:2012hj,Lee:2012bf,Siemens:2013zla}. The overlap reduction function for a pulsar pair $I$ and $J$ reads
\begin{align}
    \Gamma_{IJ} (f) = \mathcal{R}_{\rm PTA} (f) \zeta_{IJ}(\psi),
\end{align}
where $\mathcal{R}_{\rm PTA} (f) = 1/(12\pi^2 f^2)$ is the averaged timing residual response function for a single pulsar \cite{Hazboun:2019vhv,Chamberlin:2014ria}, and $\zeta_{IJ}(\psi)$ is the Hellings-Downs factor \cite{Hellings:1983fr} with a separating angle $\psi$ between two pulsars in the sky, it reads
\begin{align}
\zeta_{IJ}(\psi) = \frac{1}{2} \left[ \delta_{IJ} + 1 + c_{\psi} \left(3 \, \hbox{ln}c_\psi  - \frac{1}{2}\right) \right],
\end{align}
where $c_{\psi} \equiv (1-\cos \psi)/2$, and $\zeta_{II}=1$ for a single pulsar. Due to unknown pulsar distribution in future experiment, we used the averaged (for angle $\psi$) value $\zeta_{\rm rms} = 0.147$ \cite{Schmitz:2020syl} in our numerical. Then we average the means for each set of measurements as
\begin{align}
    \hat{\mu} = \frac{\sum\limits_{I=1}^{N_{\rm SKA}}\sum\limits_{J>I}^{N_{\rm SKA}} \lambda_{IJ} \, \mu_{IJ}}{\sum\limits_{I=1}^{N_{\rm SKA}}\sum\limits_{J>I}^{N_{\rm SKA}} \lambda_{IJ}} \simeq \mu,
\end{align}
and the corresponding averaged variance
\begin{align}
  \notag  \sigma^2_{\hat{\mu}} & \equiv \langle \hat{\mu}^2 \rangle - \langle \hat{\mu}\rangle^2 =  \frac{\sum\limits_{I=1}^{N_{\rm SKA}}\sum\limits_{J>I}^{N_{\rm SKA}} \lambda_{IJ}^2 \, \sigma_{IJ}^2}{\left(\sum\limits_{I=1}^{N_{\rm SKA}}\sum\limits_{J>I}^{N_{\rm SKA}} \lambda_{IJ}\right)^2} \\ & \simeq \frac{2}{N_{\rm SKA}(N_{\rm SKA}-1)} \sigma^2
\end{align}
where $\lambda_{IJ}$ are weighted constant, and $N_{\rm SKA}$ is a number of monitoring pulsars. The approximation is assuming that all the observation properties are the same, i.e.~identical Hellings-Down factor, $\lambda_{IJ}$, and
\begin{align}
    \mu_{IJ}\equiv \mu, \;\;\;\; \hbox{and} \;\;\;\;\; \sigma_{IJ} \equiv \sigma
\end{align}
for each pulsar pair $IJ$. Since we are considering a strong stochastic GW background, the self-noise dominates (or approximately equals to) the noise contributions in our study in SKA frequency sensitivity region (so-called strong signal regime), i.e.
\begin{align}
\label{Eq: app: S_signal gg D_noise}
    S_{\rm signal} (f) R_{\rm PTA}(f) \gg D_{\rm noise}(f).
\end{align}
In such a scenario, the variance Eq.(\ref{Eq app sigma}) for a pulsar pair with pulsar $I$ and $J$ simplifies to
\begin{align}
\notag \sigma^2_{IJ} \simeq & \; \frac{t_{\rm obs}}{2} \int^\infty_0 df\, \left[ \Gamma_I(f)\Gamma_J(f) + \Gamma_{IJ}^2(f) \right]S^2_{\rm signal}(f)\\
\simeq & \; \frac{t_{\rm obs}}{2} \int^\infty_0 df\, \Gamma_I^2(f) S^2_{\rm signal}(f)
\end{align}
where we ignore the $D_{\rm noise}(f)$ terms from Eq.(\ref{Eq app sigma}) at first approximation, and the second approximation is due to $\bar{\zeta}_{\rm rms}^2  \ll 1$. Following the steps in Eq.(\ref{Eq: app: Delta Omega gw fi}), we obtain the GW spectrum uncertainty for each frequency bin $f_i$ for pulsar timing array experiment 
\begin{align}
\notag \Delta\Omega_{\rm gw} (f_i) & = \frac{2\pi^2}{3H_0^2} \frac{f_i^3}{\Gamma_{IJ}(f_i) t_{\rm obs} \Delta f_{i}}\sigma (f_i) \\ \notag & \simeq \frac{1}{\sqrt{N_{\rm SKA}(N_{\rm SKA}-1)}} \frac{2 \pi^2} {3 H_0^2} \frac{f_i^5}{\zeta_{\rm rms} \sqrt{ t_{\rm obs} \Delta f_i}} \\ & \times \left[ \frac{1}{\Delta f_i} \int_{F_i} df \, f^{-4} S_{\rm signal}^2(f)  \right]^{1/2},
\end{align}
where we have used Eq.(\ref{Eq: app: sigma fi}). The effective GW noise spectrum in the limit $\Delta f_i \to df$ (see calculations around Eq.(\ref{Eq: app: effective Omega noise})) for a given GW spectrum $\Omega_{\rm gw}(f)$ reads,
\begin{align}
\label{Eq: app: Omega noise eff f}
    \Omega_{\rm noise}^{\rm eff}(f) \simeq \frac{\sqrt{2}}{\zeta_{\rm rms}\sqrt{N_{\rm SKA}(N_{\rm SKA}-1)}} \Omega_{\rm gw}(f).
\end{align}
The FNR estimation is given in Eq.(\ref{Eq: app: FNR}) with $n_{\rm obs}=2$ for a pulsar pair. We also include both self noise and instrumental noise in our numerical.

\subsubsection{SKA parameters}

The Square Kilometre Array (SKA) is a planned next generation intergovernmental radio telescope
\cite{Janssen:2014dka,Carilli:2004nx,Weltman:2018zrl}. It has expected to detect nano-Hertz frequency band with pulsar time array observation. We consider the operating time $t_{\rm SKA} = 20\,$yrs, cadence $t_c = 1\,$week \cite{Janssen:2014dka,Moore:2014lga}, and a constant SKA instrumental noise spectrum $D_{\rm noise}^{\rm SKA} (f) \simeq 1.1 \times 10^{-9} \, \hbox{Hz}^{-3}$ \cite{Schmitz:2020syl,Thrane:2013oya}. The sensitive frequency range is between the operating period $f_{\rm min} = 1/t_{\rm SKA}$ and the cadence of the timing observation, $f_{\rm max}=1/t_c$. As discussed in the main letter, we chase a signal that is about $1\%$ of the observation GW background, i.e.~$\Delta \Omega_{\rm gw}^{\rm th}(f)/\Omega^{\rm st}_{\rm gw}(f) \sim 1\%$ (see Eq.(\ref{Eq: app: Omega th})). In order to observe such a small signal, we need
\begin{align}
  \frac{\sqrt{2}}{\zeta_{\rm rms}\sqrt{N_{\rm SKA} (N_{\rm SKA}-1)}} \lesssim 1\%,
\end{align}
for ensuing a small self-noise spectrum from Eq.(\ref{Eq: app: Omega noise eff f}). This implies a large monitoring pulsar number,
\begin{align}
    N_{\rm SKA} \gtrsim \mathcal{O}(10^3).
\end{align}
Such a large $N_{\rm SKA}$ is an optimistic estimation (maybe unrealistically optimistic). For comparison, Ref.~\cite{Moore:2014lga} discussed that $N_{\rm SKA}\simeq 50$ is a conservative estimation, and this value has been widely quoted in literature, see e.g.~\cite{Schmitz:2020syl,Siemens:2013zla}. Moreover, Ref.\cite{Combes:2021xez} mentioned that SKA will detect more pulsars in the sky, but it is unknown how many pulsars will be monitored for high-precision GW observation.

\subsection{Auto-Correlated Detector}\label{subsection: ACD}

We analyze the statistical quantities in this subsection with targeting the auto-correlated detectors such as LISA. Follow Eq.(\ref{Eq app Dnoise}) and Eq.(\ref{Eq app sIsJ}) with considering a single detector, we have the mean $\mu$ as
\begin{align}
   \notag \mu \equiv \langle s^2 \rangle & = \int^{+\infty}_{-\infty} df\, \langle \tilde{s}(f) \tilde{s}^*(f) \rangle \\ &  = \frac{t_{\rm obs}}{2} \int^{+\infty}_{-\infty} df\, \mathcal{R}(f) S_{\rm signal}(f),
\end{align}
where $\tilde{s}(f)$ is the GW signal as given in Eq.(\ref{Eq: app: SI sI nI}), $\mathcal{R}(f)$ is the detector polarization- and sky-averaged response function, and the ensemble averaged signal production is
\begin{align}
    \langle \tilde{s}(f) \tilde{s}^*(f^\prime) \rangle = \frac{1}{2} \delta(f-f^\prime) \mathcal{R}(f) S_{\rm signal}(f).
\end{align}
Its variance is (similar to Eq.(\ref{Eq app sigma})),
\begin{align}
\notag \sigma^2 =&\, \langle (\tilde{S}(f)\tilde{S}^*(f)^2 \rangle - \langle \tilde{S}(f)\tilde{S}^*(f) \rangle^2\\
\notag =&\, \int_{-\infty}^{+\infty} df df^\prime\, \Bigg[ \langle \tilde{S}^*(f) \tilde{S}(f) \tilde{S}^*(f^\prime) \tilde{S}(f^\prime) \rangle \\ \notag & - \langle \tilde{S}^*(f) \tilde{S}(f)\rangle \langle \tilde{S}^*(f^\prime) \tilde{S}(f^\prime) \rangle \Bigg]\\
\notag =&\, \frac{t_{\rm obs}}{4} \int^{+\infty}_{-\infty} df \Bigg[ 2R^2(f) S_{\rm signal}^2(f) \\ \notag & + 2 \mathcal{R}(f) S_{\rm signal}(f) D_{\rm noise}(f) + 2 D_{\rm noise}^2(f)  \Bigg]\\ \notag =&\, t_{\rm obs}\int_0^\infty df \Bigg[ \mathcal{R}^2(f) S_{\rm signal}^2(f) \\  & + \mathcal{R}(f) S_{\rm signal}(f) D_{\rm noise}(f) + D_{\rm noise}^2(f)  \Bigg]
\end{align}
where for the noise in Eq.(\ref{Eq: app: SI sI nI}), we have production
\begin{align}
    \langle \tilde{n}(f) \tilde{n}^*(f^\prime) \rangle = \frac{1}{2} \delta(f-f^\prime) D_{\rm noise}(f).
\end{align}
Follow the procedure as given on above Eq.(\ref{Eq: app: FNR}), the corresponding uncertainty at frequency bin $\{f_i\}$ for a given GW spectrum $\Omega_{\rm gw}(f)$ is
\begin{align}
    \Delta \Omega_{\rm noise}^{\rm eff} (f_i) \equiv \frac{2\pi^2}{3H_0^2} \frac{f_i^3}{\mathcal{R}(f_i) \sqrt{ t_{\rm obs} \Delta f_i}}  \sqrt{ \bar{A}_{\rm auto}(f_i) }
\end{align}
with
\begin{align}
    \notag A_{\rm auto}(f) \equiv & \mathcal{R}^2(f) S_{\rm signal}^2(f) \\ & + \mathcal{R}(f)S_{\rm signal}(f) D_{\rm noise}(f) + D_{\rm noise}^2(f),
\end{align}
and
\begin{align}
   \bar{A}_{\rm auto}(f_i) \equiv \frac{1}{\Delta f_i} \int_{F_i} df\, A_{\rm auto}(f),
\end{align}
where $F_i = [ f_i - \Delta f_i/2, f_i + \Delta f_i/2]$. Comparing the signal difference as in Eq.(\ref{Eq: app: Omega th}) with a limitation $\Delta f_i\to df$, we obtain the effective GW noise spectrum
\begin{align}
    \Omega^{\rm eff}_{\rm noise} (f) = \frac{2 \pi^2}{3 H_0^2} f^3 \sqrt{\frac{A_{\rm auto}(f)}{\mathcal{R}^2(f)}},
\end{align}
and the fluctuation-to-noise ratio (FNR) for the detectability of the given GW spectrum difference $\Delta \Omega^{\rm th}_{\rm gw}(f)$ in auto-correlated detection,
\begin{align}
    \hbox{FNR} \equiv \left[ t_{\rm obs} \int df\, \left( \frac{\Delta \Omega_{\rm gw}^{\rm th}(f)}{\Omega^{\rm eff}_{\rm noise}(f)}\right)^2\right]^{1/2}.
\end{align}

\subsubsection{LISA parameters}

LISA experiment parameters can be found in LISA 2017 mission proposal \cite{LISA:2017pwj}, here we introduce the functions that we used in our numerical. The LISA response function estimates to \cite{Caprini:2019pxz}
\begin{align}
    \notag \mathcal{R}(f) \simeq & 16 \sin^2\left( \frac{2 \pi f L_{\rm LISA}}{c_{\rm light}} \right) \frac{3}{10} \\ & \times \frac{1}{1+0.6 \left( 2\pi f L_{\rm LISA}/c_{\rm light} \right)^2}   \left( \frac{2 \pi f L_{\rm LISA}}{c_{\rm light}} \right)^2,
\end{align}
where $L_{\rm LISA} = 2.5 \times 10^{9} \,$m is the LISA arm length, and we assume all the LISA arms are equal and constant. LISA instrumental noise are from a single mass acceleration noise
\begin{align}
 \notag    D_{\rm acc}(f) = & \,\left( A \frac{\hbox{fm}}{\hbox{s}^2 \sqrt{\hbox{Hz}}} \right)^2 16\sin^2\left( \frac{2 \pi f L_{\rm LISA}}{c_{\rm light}} \right) \\ \notag & \times \left[ 3 + \cos^2 \left( \frac{4 \pi f L_{\rm LISA}}{c_{\rm light}} \right) \right]  \left[ 1 + \left( \frac{0.4 \, \hbox{mHz}}{f} \right)^2 \right] \\ & \times \left[ 1 + \left( \frac{f}{8\,\hbox{mHz}} \right)^4 \right] \left( \frac{1}{2\pi f} \right)^4 \left( \frac{2 \pi f}{ c_{\rm light}} \right)^2,
\end{align}
and the optical metrology system noise,
\begin{align}
  \notag  D_{\rm oms}(f) = & \left( P_A \frac{\hbox{pm}}{\sqrt{\hbox{Hz}}} \right)^2 16 \sin^2 \left( \frac{2 \pi f L_{\rm LISA}}{c_{\rm light}} \right) \\ & \times \left[ 1+ \left( \frac{2 \, \hbox{mHz}}{f} \right)^4 \right] \left( \frac{2 \pi f}{c_{\rm light}}  \right)^2,
\end{align}
where we adopt the acceleration $A$ and optical $P_A$ amplitude parameters as \cite{LISA:2017pwj}
\begin{align}
    A = 3,\;\;\;\; P_A = 10.
\end{align}
The total noise spectrum for LISA would be
\begin{align}
    D_{\rm noise}(f) = D_{\rm acc}(f) + D_{\rm oms}(f).
\end{align}
We also adopt the extended LISA observation period $t_{\rm obs} = 10\,$years, and the sensitive frequencies $20 \mu \,\hbox{Hz}\leq f \leq 1\,$Hz.

\bibliographystyle{apsrev4-1}
\bibliography{ArXiv}

\end{document}